\documentclass[aps,pra,twocolumn,showpacs,showkeys,groupedaddress]{revtex4}

\usepackage{amsmath}
\usepackage{amssymb}
\usepackage{mathrsfs}
\usepackage{graphicx}
\usepackage{textcomp}
\usepackage{color}
\begin{document}

\title{Diffraction-limited Fabry-P\'{e}rot Cavity in the Near Concentric Regime}

\author{Kadir Durak$^{1}$, Chi Huan Nguyen$^{1}$, Victor Leong$^{1}$, Stanislav Straupe$^{1,2}$,\\ Christian Kurtsiefer$^1$ }
\affiliation{$^1$Centre for Quantum Technologies, National University of Singapore, Singapore\\
            $^2$Faculty of Physics, M.V.Lomonosov Moscow State University, Moscow, Russia}

\date{\today}

\begin{abstract}
Nearly concentric optical cavities can be used to prepare optical fields with a
very small mode volume. We implement an anaclastic design of a such a cavity
that significantly simplifies mode matching to the fundamental cavity
mode. The cavity is shown to have diffraction-limited performance for a mode
volume of $\approx10^4\lambda^3$. This is in sharp contrast with the behavior of
cavities with plano-concave mirrors, where aberrations significantly increase
the losses in the fundamental mode. We estimate the related cavity QED
parameters and show that the proposed cavity design allows for strong coupling
without a need for high finesse or small physical cavity volume.
\end{abstract}

\maketitle

\section{Introduction}

Achieving strong interaction of single quantum emitters with electromagnetic field in a single-photon regime is one of the ever-sought goals in modern atomic physics. Besides fundamental interest it is motivated by needs of quantum information science, where information exchange between ``flying qubits'' encoded in photonic degrees of freedom and ``stationary qubits'' realized in the atomic or other microscopic material systems lies in the heart of various communication protocols and computational architectures \cite{Kimble_Nature_2008}.

One of the well established approaches to achieve the desired coupling is to
enhance photon-atom interaction in high-finesse cavities
\cite{Kimble_PhisScripta_1998}. Since the early demonstrations
\cite{Ye_PRL_1999} the field of cavity QED with single atoms was a constant
struggle for higher coupling \cite{Boozer_PRL_2007, Wilk_Science_2007,
  Ritter_Nature_2012} mostly relying on ultra-high-reflectivity coatings of
constantly increasing sophistication \cite{Hood_PRA_2001}. At the same time
the mode volume of a cavity aiming at strong coupling must be kept as small as
possible, which usually results in some sort of a microresonator, be it a
micro Fabry-P\'{e}rot cavity \cite{Steinmetz_APL_2006} or some kind of a
monolithic whispering gallery resonator \cite{Vahala_Nature_2003}. Recently,
also photonic waveguide structures have been successfully used to achieve this
goal \cite{Akahane2003, Yoshie2004, Sato2011}.

An alternative route to small mode volume is to use the strongly focused
``hourglass modes'' of near-concentric cavities \cite{Morin_PRL_1994,
  Daul_EPJD_2005, Haase_OptLett_2006}. Here we follow this route and
demonstrate an effective coupling of light to a Fabry-P\'{e}rot resonator near
the stability limit. Mode matching of the external Gaussian beam to such a
cavity is problematic and we provide arguments, both experimental and
numerical, that optical aberrations in the mirrors are one of the main reasons
of these problems. A cavity mirror design, initially proposed in
\cite{Aljunid_JMO_2011} is experimentally tested and shown to be superior
over traditional mirror geometries. The paper is organized as follows: we
begin with demonstrating the problems of conventional mirrors in concentric
cavities in Section~\ref{plano-concave}, analyze their origins numerically in
Section~\ref{aberrations}, describe the cavity lens design and its experimental test in Section~\ref{design}, and estimate the expected coupling to single atoms in Section~\ref{coupling}.

\section{Concentric cavity with plano-concave mirrors \label{plano-concave}}

The small mode volume optical cavity with the length approaching the concentric point makes it extremely vulnerable to various instabilities. Our first goal was to study the behavior of an ``ordinary'' cavity under these extreme conditions. The cavity was formed by two mirrors on a plano-concave substrate of BK-7 glass. The planar side had anti-reflection coating at 780 nm, while the spherical surface with 50~mm radius of curvature was coated for 0.978 reflectivity at the same wavelength.

\begin{figure}
  \begin{center}
    \includegraphics[width=0.8\columnwidth]{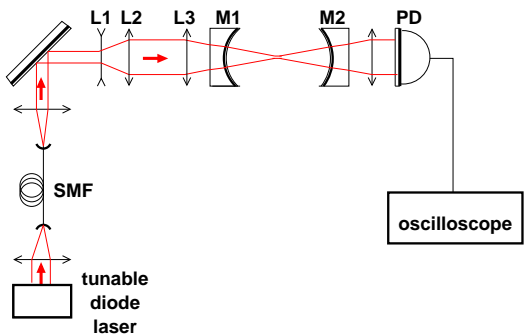}
    \caption{
    Experimental setup with a test cavity formed by two plano-concave mirrors
    M1,M2 with 0.978 reflectivity, 50\,mm radius of curvature and 6.35\,mm
    aperture. L1, L2, L3 - spatial mode-matching optics, SMF- single mode fiber for 780 nm beam and PD - photodiode.
    \label{exp_setup}}
  \end{center}
\end{figure}

The experimental set up used to determine the cavity parameters is shown in
Fig.~\ref{exp_setup}. We used an extended cavity diode laser with wavelength
of 780 nm as a light source with tunable frequency. The laser beam was
spatially mode-cleaned by a single-mode fiber, and mode matched to a cavity
with a three lens system. The transmitted light was detected by a photodiode
and recorded. Part of the probe light was sent to a rubidium reference cell
(not shown) where a Doppler-free spectroscopy signature in Rubidium provided
an absolute frequency reference.

In this transmission experiment it is more convenient to scan the laser
frequency by means of a diffraction grating, rather than scanning the cavity
length. This is because for a cavity very close to the concentric
configuration, the variation of the mirror separation on the order of half a
wavelength in order to observe one free spectral range (FSR) significantly
changes the transverse mode. This in turn would require an adjustment of the
mode-matching optics during the length variation. The mode matching components
L1, L2 and L3 are chosen and positioned accordingly for each time we change the
cavity length.

\begin{figure}
\begin{center}
    \includegraphics[width=0.8\columnwidth]{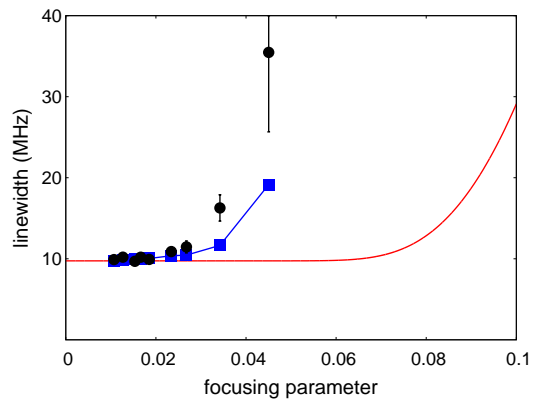}
    \caption{Linewidth of a cavity formed by plano-concave mirrors, measured 
      for different focusing parameters $u$ (circles). The solid line
      corresponds to a simple model taking into account only diffraction
      losses due to finite size of the mirrors, while the squares represent
      calculations considering both aberrations and diffraction losses (the
      joining life is added to guide the eye only).\label{test_cavity}}
\end{center}
\end{figure}

Figure~\ref{test_cavity} shows the observed cavity transmission linewidth as a
function of the focusing parameter $u=2w/L$, defined as a ratio of
the input beam waist at the cavity mirrors $w$ to half of the cavity length
$L$.  We will discuss several quantities of interest versus this dimensionless
focusing parameter $u$ instead of cavity length to allow for direct
comparison of the results for different cavities. Almost all significant
changes in behavior are observed within few micrometers from the concentric
length $L=2R$, where $R$ is the radius of curvature of the mirrors. The
linewidth increases dramatically as the cavity length approaches the
concentric limit, implying increasing losses for the fundamental cavity
mode. Partially that can be explained by increase in mode waist at the mirrors
leading to high losses due to finite aperture of the mirrors. These losses
(which we will refer to as \emph{diffraction losses}) for a fundamental
Gaussian mode can be approximately taken into account by introducing a
correction of the fraction of power left in the cavity after one
round trip
\begin{equation}
    \label{losses}
    \rho=\rho_0\left(1-\exp\left[-\frac{2a^{2}}{w^{2}}\right]\right)^{2},
\end{equation}
where $\rho_0$ is the squared reflectivity of the mirrors (assuming it is same
for both mirrors), $a$ the radius of the mirror aperture, and $w$ is the
cavity mode waist at the mirror. The resulting cavity finesse
\begin{equation}
    \label{finesse}
    \text{\ensuremath{\mathcal{F}\left(\rho\right)}=}\frac{\pi}{2\arcsin\left(\frac{1-\sqrt{\rho}}{2\sqrt[4]{\rho}}\right)}
\end{equation}
leads to a linewidth $\kappa=c/\left(2L\mathcal{F}\right)$, assuming only
diffraction losses. Fig.~\ref{test_cavity} shows this estimation as a
solid line. However, the expression significantly underestimates the
measured values. Even though (\ref{losses}) is only an approximation and exact
calculation of diffraction losses requires numerical solution of the
diffraction equation \cite{Fox_Li_IEEE_1966}, the approximation should be
valid for our purposes since the fundamental mode waist at the cavity mirrors
in the region of interest is significantly smaller than the mirror
aperture. We therefore explore aberrations as another explanation for the
observed behavior.

\section{Aberration analysis\label{aberrations}}

At the concentric limit the waist of a cavity mode is almost at the diffraction limit and the input beam has to be strongly focused to match it. Some amount of optical aberrations will be inevitably introduced by spherical mode-matching optics, and most importantly by the planar surface of the input mirror itself. Aberrations degrade the Gaussian input mode and cause significant coupling to higher order spatial modes of the cavity.

For a cavity with cylindrical symmetry, a suitable set of spatial modes is described (in dimensionless units) by Laguerre-Gaussian functions:
\begin{eqnarray}
    &\Psi_{l,p}\left(r,\phi,z\right)  =  \frac{C_{l,p}}{w(z)}\left(\frac{r\sqrt{2}}{w\left(z\right)}\right)^{\left|l\right|}\exp\left[-\frac{r^2}{w\left(z\right)^2}\right]\nonumber\\
    &    \times L_{p}^{\left|l\right|}\left(\frac{2r^2}{w\left(z\right)^2}\right)
      \times \exp\left[ik\frac{r^{2}}{2R(z)}\right]\exp\left[il\phi\right]\\
    &   \times \exp\left[-i\left(2p+\left|l\right|+1\right)\xi\left(z\right)\right],\nonumber\\
    \nonumber
\end{eqnarray}
where $L_{p}^{\left|l\right|}$ are generalized Laguerre polynomials, $r$ is
the transverse distance from the optical axis, $w\left(z\right)$ the mode
waist at position $z$, $p$ the radial mode number, $l$ the azimuthal index
with $\left|l\right|\leq p$, $R(z)$ the radius of curvature of the wavefront
at $z$, $\xi\left(z\right)=\arctan(z/z_{R})$ the longitudinal Guoy phase, and
$z_{R}$ the Rayleigh range. A normalization constant $C_{l,p}$ ensures
$\int\left|\Psi_{l,p}\left(r,\phi,z_m\right)\right|^{2}rdrd\phi=1$ at the 
mirror position $z_{m}$.

The frequency shift of the higher order modes ($p,l>0$) with respect to the
fundamental one ($p,l=0$) is given by
\begin{equation}
\Delta\nu_{l,p}=\frac{c}{2\pi L}\left(|l|+2p\right)\mathrm{arccos}\left(1-\frac{L}{R}\right)\,,
\end{equation}
where
$c$ is the speed of light, $L$ the
cavity length, and $R$ the radius of curvature of the mirrors. In the limit
$L\approx 2R$ the mode separation becomes equal to the free spectral range of
the cavity. Higher order spatial modes then overlap, and it becomes impossible to resolve them in the frequency domain. This overlapping of modes results in the broadening of the transmission peak if the cavity length is very close to the concentric configuration.

\begin{figure}
    \includegraphics[width=0.44\columnwidth]{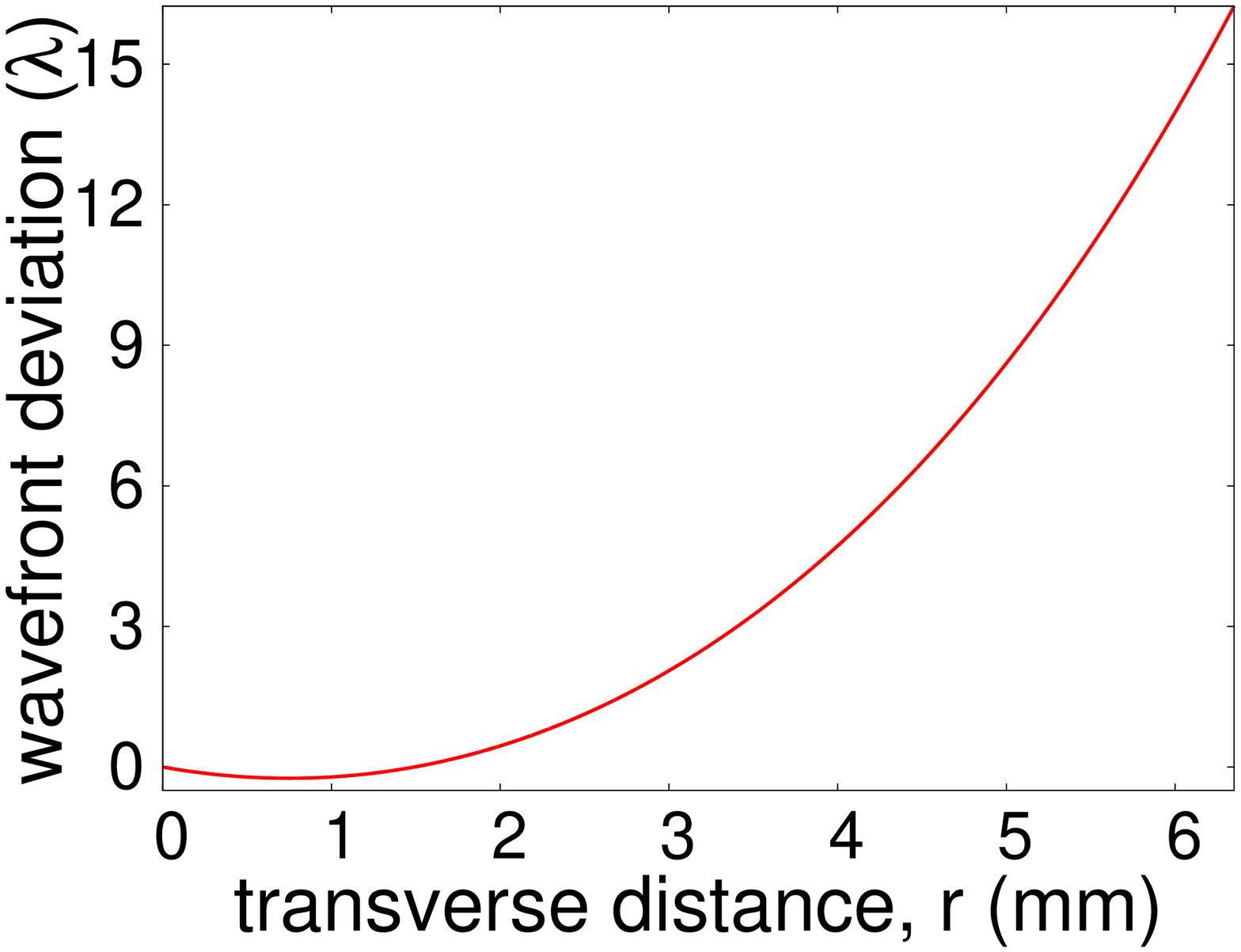}
    \includegraphics[width=0.46\columnwidth]{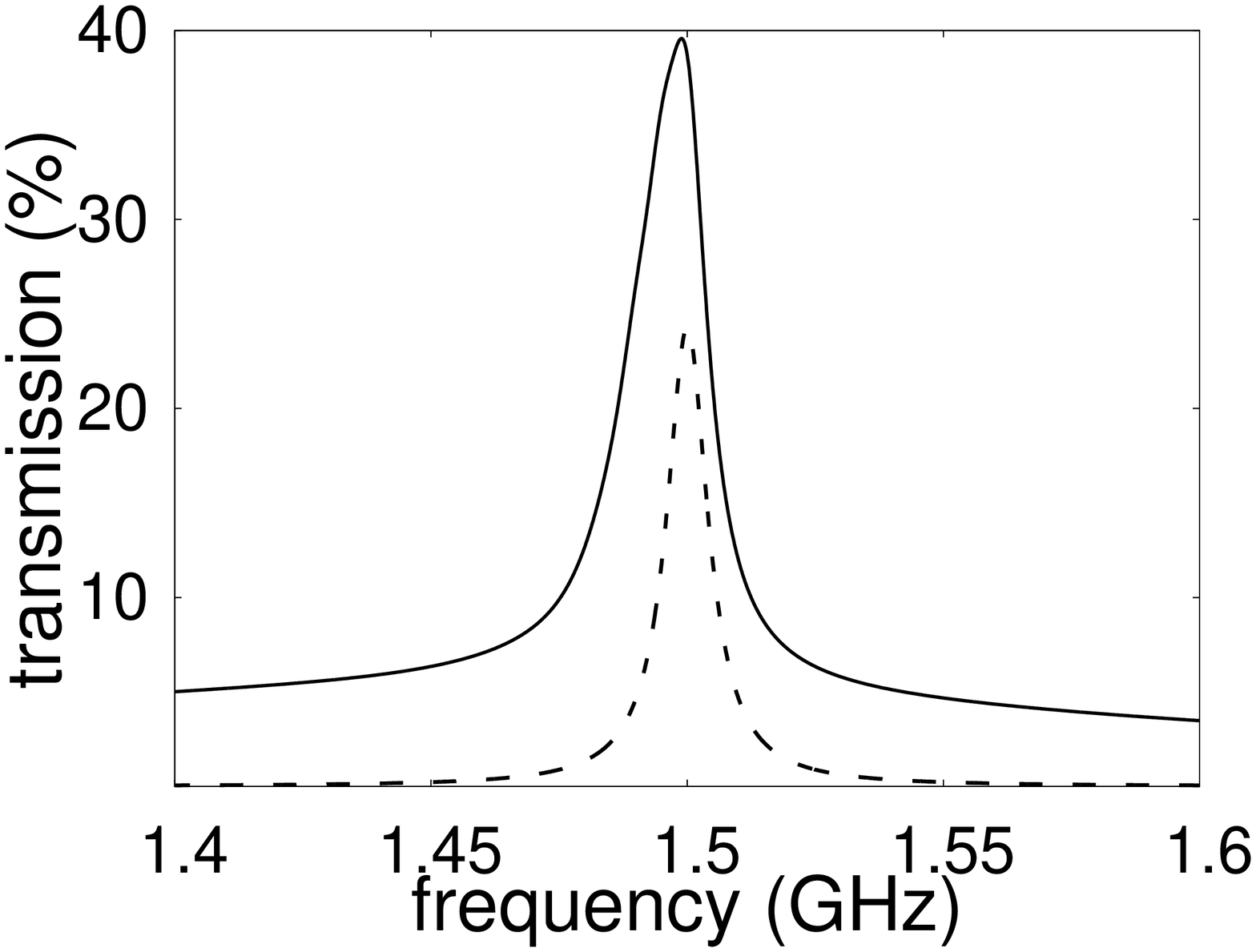}
    \caption{Aberrations in plano-concave cavity. (Left) Deviation of the
      wavefront of the input mode from the spherical shape of the mirror
      surface as a function of transverse distance from the optical axis (in
      fractions of wavelength). (Right) Simulated transmission spectrum for
      the superposition of cavity modes excited by an aberrated beam (solid
      line) compared to the fundamental mode of the cavity (dashed line). Both
      figures correspond to the value of focusing parameter $u=0.047$.
    \label{deviation}}
\end{figure}

To make these considerations quantitative, we determined the coupling of the aberrated input beam that we used in the measurement to higher order spatial modes of the cavity numerically. The wavefront deformation of the input beam at the surface of the input mirror was estimated by ray tracing. By following the optical path including all the mode matching optics, we determine the phase of the input beam at the spherical surface of the mirror with respect to the transverse distance from the optical axis. The phase of the beam at the optical axis is taken as the phase reference. Figure~\ref{deviation} shows the wavefront deviation from the ideal spherical wavefront of the Gaussian cavity mode for a 2.35~mm input beam waist at mirror position (5.7~$\mu$m waist in the center of the cavity), corresponding to $u=0.047$.

Assuming that mode-matching is affected by this wavefront distortion only, we
can calculate the coupling coefficients. We express the spatial mode of the
input beam as a fundamental Gaussian mode of the cavity, multiplied by a
slowly varying  complex phase term:
\begin{eqnarray}
    \label{input_mode}
    &\xi\left(r,\phi,z\right) = \frac{C}{\omega(z)}\exp\left[-\frac{r^2}{w\left(z\right)^2}\right]\nonumber\\
    &\times\exp\left[ik\frac{r^{2}}{2R(z)}\right]\exp\left[-i\xi\left(z\right)\right] \times\exp\left[i\varphi\left(r\right)\right],
\end{eqnarray}
where $\varphi\left(r\right)$ is the calculated phase retardance of the input
beam with respect to the cavity mode. The coupling of the input beam to a
spatial mode $\Psi_{l,p}$ can be characterized by a normalized intensity
$\gamma_{l,p}$ of the corresponding mode excited by the input beam in a
spatial mode $\xi(r,\phi,z)$, which is given by the squared modulus of an
overlap integral:
\begin{equation}
    \label{overlap}
    \gamma_{l,p}=\left|\int\limits_{0}^{a}\int\limits_{0}^{2\pi}\Psi_{l,p}(r,\phi,z_m)^{*}\xi(r,\phi,z_m) rdrd\phi\right|^2\,,
\end{equation}
taken at $z=z_m$ corresponding to the input mirror position. The finesse (and
linewidth) for higher order modes can be evaluated expression (\ref{finesse}),
but with different diffraction losses per round trip taken into for each mode
($l,p$):
\begin{equation}
    \rho_{l,p}=\rho_0\left(\int\limits_{0}^{a}\int\limits_{0}^{2\pi}\left|\Psi_{l,p}(r,\phi,z_m)\right|^2 rdrd\phi\right)^2
\end{equation}

In the case of $\varphi(r)\equiv 0$ only the fundamental Gaussian mode has
non-zero overlap with the input mode, while for an aberrated beam
(\ref{input_mode}) higher order modes are significantly populated. For every
experimental point in Fig.~\ref{test_cavity}, the mode populations were
calculated numerically, including modes up to $p=50$. The transmission
spectrum was calculated as a superposition of transmission lines for each mode
with maxima shifted by $\Delta\nu_{l,p}$ and line width
$\kappa_{l,p}=c/\left(2L\mathcal{F}_{l,p}\right)$. An example of the
calculated spectrum for the maximal experimentally achieved focusing parameter
of $u=0.047$ is shown in Fig.~\ref{deviation}. We took the full width at
half-maximum of this spectrum as an estimate of the experimentally observed
linewidth, calculated values are shown in Fig.~\ref{test_cavity} along with
experimental data. The error on the measured linewidth here is the standard
deviation of the full width at half-maximum over 100 sweeps. The error on the
focusing parameter is evaluated through the mode waist at the center of the
cavity, which is found by measuring the error of the minimum waist at the
optical axis at one single pass of the beam (absent second mirror). The error
of the focusing parameter is less than 2\% of the focusing parameter values
for Fig.~\ref{test_cavity}.
One can observe reasonable correspondence between the experimental data and the numerical simulation results, supporting our hypothesis about the major role of aberrations in mode-matching for the cavity in near-concentric configuration.

In this analysis, there are two basic assumptions made. First, we assume that
the input mode through the first substrate surface to the mirror surface can be
approximated via a ray tracing method. This seems justified because the radius
of curvature of the wavefronts there is much larger than the optical
wavelength. Second,
Laguerre-Gaussian modes are taken as the cavity eigenmodes. However, the
cavity mirrors have a finite size, and a numerical calculation of real
cavity eigenmodes is required. An example of this treatment can be found in
\cite{Kleckner_PRA_2010}. However, even in our experimentally accessible
configuration closest to concentric case, the mode waist at the mirror is
smaller than the radius of the mirror aperture (0.42 mirror radius). It can be
seen on Fig.~\ref{test_cavity} that the diffraction loss is not significant
even at the closest to concentric configuration data point with a focusing
parameter of 0.047.

Thus, the use of Laguerre-Gaussian modes as the cavity eigenmodes is a
reasonable approximation. In other words, the linewidth broadening within the
near concentric regime is due to the population of the higher order modes,
which is an obstacle to observe the mode splitting because of the cavity-atom
interaction. Our aberration analysis of the near concentric cavity regime
therefore suggests that in order to observe the cavity quantum electrodynamic effects, one needs to avoid the aberrations of the input mode.

\section{Anaclastic cavity design\label{design}}
One way to eliminate the aberrations of the input mode is to use anaclastic
design of mode-matching optics. An anaclastic lens has an aspheric surface
converting the plane wavefront of a collimated Gaussian input beam to a
converging spherical wavefront. A design of cavity mirrors incorporating such
an aspheric surface as the input surface of the cavity mirror was proposed in
\cite{Aljunid_JMO_2011}, but has in fact been known for a very long time \cite{ibnisahl, 1990}.
\begin{figure}
  \begin{center}
    \includegraphics[width=0.5\columnwidth]{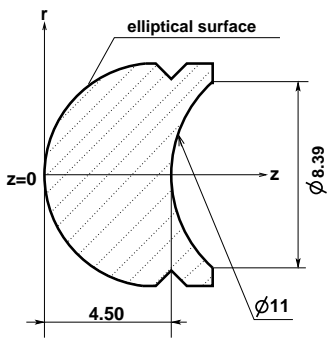}
    \end{center}
    \caption{Cross section of the anaclastic cavity lens design. The
      aspherical surface is an ellipsoid of revolution defined by $\left(1-z/a\right)^2-\left(r/b\right)^2=1$, with half-axes $a=6.3844$\,mm and $b=5.2620$\,mm. This surface acts as a lens with a focal point at $z=10$\,mm.
    \label{anaclastic}}
\end{figure}
The aspheric surface is an ellipsoid of revolution with half axes $a=fn/(n+1)$ in longitudinal and $b=f\sqrt{(n-1)/(n+1)}$ in transverse direction, where $f$ is the desired focal length, and $n$ is the refractive index of the material used. If the second spherical surface is centered at the focus of the lens, it does not introduce any distortions to the wavefront of the input beam resulting in an aberration-free design. The drawing of the cavity mirror used in this work is shown in Fig.~\ref{anaclastic}. The mirrors were made of N-SF11 glass with refractive index of $n=1.76583$ at 780\,nm with the focal length $f=10$\,mm, corresponding to 5.5\,mm radius of curvature of the spherical cavity mirrors. The elliptical surface was anti-reflection coated for 780\,nm wavelength, and the spherical surface had a high-reflectivity coating the transmission of which was specified to be larger than $0.99$ by the manufacturer. However in what follows we use the value of 0.9936 estimated from the measured linewidth for small input beam waists, where the diffraction and (possible) aberrative losses are insignificant.

The design combining cavity mirror and mode-matching lens not only eliminates
the aberrations of the coupling optics, but also significantly simplifies
alignment, which is a major advantage for the technically challenging confocal
configuration. With expressions in \cite{Aljunid_JMO_2011} for the field
quantization, we can associate an effective mode volume for this (standing wave)
cavity of\begin{equation}
V_\text{eff}=\frac{3\lambda^2L}{4\pi R_{sc}(u)}\,.
\end{equation}
For our design value $u=0.365$, we get a value of
$V_\text{eff}\approx10^4\lambda^3$. With this particular value of focusing
parameter, the cavity--single atom cooperativity has a maximum value of
150. However, we can experimentally realize cavity mode volume as small as
$\approx4100\lambda^3$ with $u=0.73$, but at very small values of cavity mode
volume (large focusing parameter) the diffraction loss becomes
significant. This results in the broadening of the linewidth of the
transmission peak. Consequently, the cavity decay rate becomes larger, and the
cavity-single atom cooperativity decreases (comparing to the case where
$u=0.365$).  The mode-matching is achieved by simply choosing an appropriate
waist of the collimated input beam.

\begin{figure}
  \begin{center}
    \includegraphics[width=0.8\columnwidth]{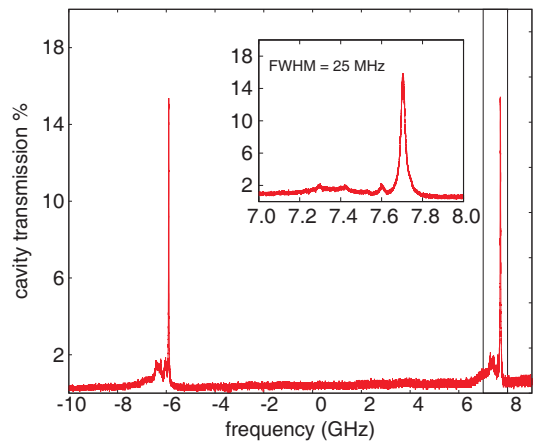}
    \caption{Transmission through an anaclastic cavity for the cavity length
      2.4~$\mu$m away from the concentric configuration, and an input
      beam waist $w=1.1.$\,mm corresponding to a focusing parameter
      $u=0.113$.\label{cavity_transmission}}
  \end{center}
\end{figure}

The performance of the anaclastic cavity design was tested in a setup similar
to those of Fig.~\ref{exp_setup}, with a three lens system replaced by a
telescope. The achieved quality of mode-matching is illustrated by
Fig.~\ref{cavity_transmission}, where a single oscilloscope trace
corresponding to a frequency scan of more than 18 GHz is shown (the cavity
free spectral range is 13.6 GHz). Some residual excitation of higher order
modes is visible, which we attribute to possibly non-ideal quality of the
aspheric surface, as well as to mismatch in input beam waist and non-perfect
beam alignment in experiment. Mode-hop-free tuning of an external cavity
diode laser over this range was accomplished by synchronizing the rotation of
the grating with adjustment of the diode current, resulting in continuous tuning
over more than 30\,GHz.

\begin{figure}
  \begin{center}
    \includegraphics[width=0.8\columnwidth]{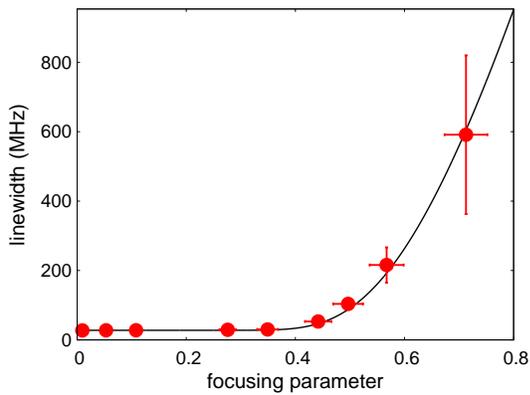}
    \caption{Measured transmission linewidth of the anaclastic cavity for
      different focusing parameter $u$ (circles). The solid line
      represents a simulation taking into account diffraction loss only.
    \label{cl_linewidth}}
  \end{center}
\end{figure}

Figure~\ref{cl_linewidth} shows the linewidth dependence on the focusing
parameter for the anaclastic cavity. In contrast to a cavity formed by
plano-concave mirrors, the linewidth broadening of the anaclastic cavity can
be completely attributed to diffraction losses. The theoretical curve in
Fig.~\ref{cl_linewidth} is calculated according to equation (\ref{finesse})
without any additional assumptions. This allows us to couple 54\% of the
cavity output power into a single mode fiber. This observation is an argument
in support of significant reduction of aberrations in the anaclastic design
even for relatively strong focusing.

\section{Estimation of single atom coupling strength\label{coupling}}

The ultimate goal of designing a small mode volume cavities is achieving a strong interaction between an electromagnetic field of the cavity mode and resonant atoms. A standard figure of merit characterizing the interaction strength is the single atom cooperativity $C=g_0^2/(\kappa\gamma)$, where $g_0$ is the coupling strength, $\kappa$ is the cavity linewidth and $\gamma$ the atomic spontaneous decay rate. As shown in \cite{Tey_NJP_2009}, the coupling strength may be expressed as
\begin{equation}
    \label{g0}
    g_0=\sqrt{\frac{\pi\gamma cR_{sc}(u)}{L}},
\end{equation}
with $L$ being the cavity length, $c$ the speed of light in vacuum, and
$R_{sc}(u)$ the dimensionless quantity characterizing the scattering
probability of a photon from the atom depending on the focusing parameter $u$,
\begin{equation}
    \label{scattering_ratio}
    R_{sc}(u)=\frac{3}{4u^3}e^{2/u^2}\left[\Gamma(-\frac{1}{4},\frac{1}{u^2})+u\Gamma(\frac{1}{4},\frac{1}{u^2})\right]^2\,.
\end{equation}

\begin{figure}
    \begin{center}
        \includegraphics[width=0.8\columnwidth]{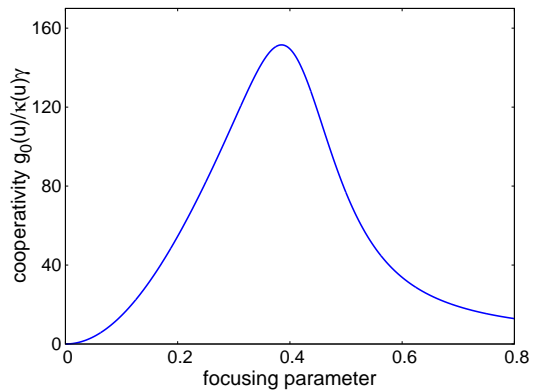}
   \end{center}
    \caption{Estimated single atom cooperativity as a function of the focusing parameter for the anaclastic cavity coupled to a D2 transition in $^{87}\mathrm{Rb}$.
    \label{cooperativity}}
\end{figure}

The estimated cooperativity for the D2 transition in $^{87}\mathrm{Rb}$ with $\gamma = 2\pi\times 6.067$~MHz and the linewidth data for the anaclastic cavity from Fig.~\ref{cl_linewidth} is shown in Fig.~\ref{cooperativity}.
The tradeoff between the increase in the scattering rate due to strong
focusing and the reduction of the cavity finesse due to higher diffraction
losses for larger beams result in an optimum value of the input beam waist
(and hence the cavity length in our design). The estimated cooperativity
reaches the maximal value of $C\approx150$, which clearly indicates that
strong coupling regime should be achievable with the presented cavity design.

\section{Conclusion}
In summary, we have studied the linewidth broadening effects in optical
cavities near the concentric limit. Optical aberrations of the input beam were
identified as a main reason for the observed broadening and numerical modeling
was performed to estimate the linewidth for the experimental data of a cavity
with plano-concave mirrors. The numerical results are in reasonable
correspondence with the experimental data supporting our claim of the
aberrative nature of the observed behavior. Our results suggest that simply
using the aberration-corrected external coupling optics will not solve the
problem, since the main source of the phase distortion for the input mode is a
planar mirror surface itself. To deal with this problem we have introduced an
aspheric design of a coupling surface to the cavity mirror, incorporating the
aberration-free coupling lens and a highly-reflective mirror in one piece. The
experimental test of a cavity with such an anaclastic design of the mirrors
has shown, that it significantly outperforms ordinary plano-concave mirrors
near the concentric limit. We were able to demonstrate significantly reduced
coupling of the input beam to higher-order spatial modes while still keeping
the coupling to the fundamental mode relatively high. An estimation of the
single atom cooperativity for the measured cavity linewidth suggests the
possibility of achieving strong coupling of the cavity mode to a single atom.

We believe the proposed cavity design to be an interesting alternative to small-volume cavities with ultra-high-reflectivity coatings dominating the field of cavity QED at present. For example, large distance between the mirrors and at the same time small volume of the ``hour-glass'' mode in the center of the cavity may be crucial for experiments with trapped ions, allowing to place the trap electrodes inside the cavity. Another major advantage of the proposed design is that there is no need for sophisticated dielectric coatings (the results reported here were obtained with a 99.36\% reflectivity coating), which significantly reduces the cost of the mirrors.

\begin{acknowledgements}
This work is supported by the National Research Foundation \& Ministry of
Education, Singapore.
\end{acknowledgements}


\end{document}